\documentclass[apl,reprint,notitlepage]{revtex4-2}
\usepackage{amsmath,amssymb}
\usepackage{mathastext}
\usepackage{graphicx}
\usepackage{dcolumn}
\usepackage{bm}
\usepackage{anysize}
\usepackage{epstopdf}
\usepackage{epsfig}
\usepackage[outercaption]{sidecap}
\usepackage[pass,letterpaper]{geometry}
\usepackage{color,colortbl} 
\usepackage{tabu}
\usepackage{boxhandler}
\usepackage{booktabs}
\usepackage[table,x11names,dvipsnames,table]{xcolor}

\newcommand{\tbcu}{W.m$^{-2}$.K$^{-1}$}

\graphicspath{{./Figures/FINAL/}}

\begin{document}
	
	\title{Machine learning enables robust prediction of thermal boundary conductance of 2D substrate interfaces}
	\author{$^1$Cameron Foss and $^{1,2}$Zlatan Aksamija}
	\affiliation{$^1$Electrical and Computer Engineering, University of Massachusetts Amherst, Amherst, MA 01002, USA \\ 
		$^2$Materials Science and Engineering, University of Utah, Salt Lake City, UT 84112, USA}
	\email{zlatan.aksamija@utah.edu}
	
\begin{abstract}
	 Two-dimensional van der Waals (vdW) materials exhibit a broad palette of unique and superlative properties, including high electrical and thermal conductivities, paired with the ability to exfoliate or grow and transfer single layers onto a variety of substrates thanks to the relatively weak vdW interlayer bonding. However, the same vdW bonds also lead to relatively low thermal boundary conductance (TBC) between the 2D layer and its 3D substrate, which is the main pathway for heat removal and thermal management in devices, leading to a potential thermal bottleneck and dissipation-driven performance degradation. Here we use first-principles phonon dispersion with our 2D-3D Boltzmann phonon transport model to compute the TBC of 156 unique 2D/3D interface pairs, many of which are not available in the literature. We then employ machine learning (ML) to develop streamlined predictive models, of which Neural Network and Gaussian process display the highest predictive accuracy (RMSE $<$ 5 M\tbcu~ and $R^2>$0.99) on the complete descriptor set. Then we perform sensitivity analysis to identify the most impactful descriptors, consisting of the vdW spring coupling constant, 2D thermal conductivity, ZA phonon bandwidth, the ZA phonon resonance gap, and the frequency of the first van Hove singularity or Boson peak. On that reduced set, we find that a decision-tree algorithm can make accurate predictions (RMSE $<$ 20 M\tbcu~ and $R^2>$0.9) on materials it has not been trained on by performing a transferability analysis. Our model allows optimal selection of 2D-substrate pairings to maximize heat transfer and will improve thermal management in future 2D nanoelectronics.
\end{abstract}

\maketitle	
	
\begin{figure*}[t]
	\includegraphics[width=.9\textwidth]{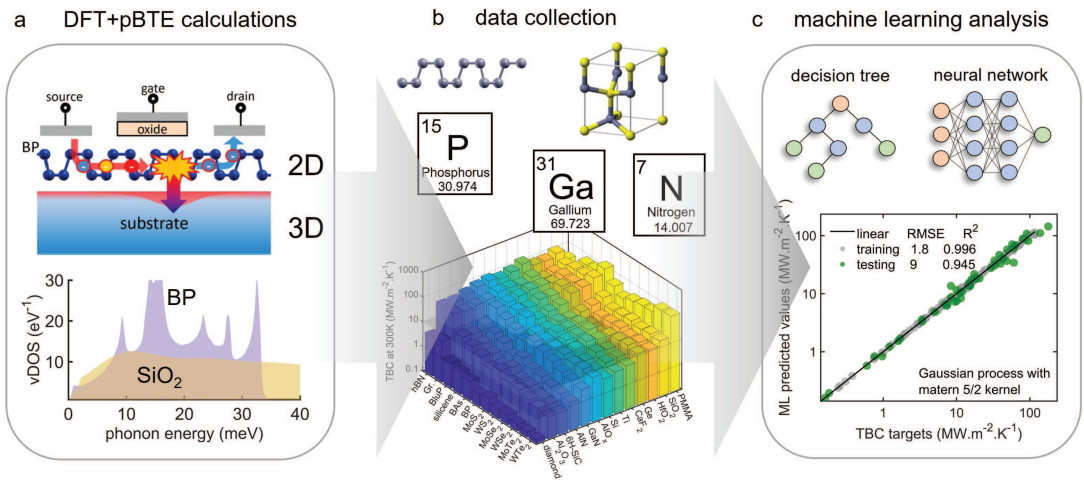}
	\caption{A schematic describing the process of data generation and processing which combines (a) rigorous atomistic simulations that are based on first-principles, (b) data collection and material descriptor selection, and (c) machine learning analysis.}\label{fig:introschem}
\end{figure*}

Two-dimensional (2D) van der Waals materials represent the limit of thinness and have potential to be breakthrough building-blocks in designing next-generation solid-state devices \cite{JariwalaACSNano14, SchwierzNano15, ZavabetiNML2020}. The weak van der Waals (vdW) bonding in their through-plane direction allows them to be exfoliated and placed on various substrates with less concern about lattice incommensurability than covalently bonded interfaces \cite{SuryavanshiJAP2019}. For the same reason, these materials exhibit high anisotropy between their covalently-bonded in-plane and vdW-bonded through-plane thermal conductivities in their bulk form \cite{QianAPL18,Chen2DMats2019,ChenFM2020,KimNature2021}. While this anisotropy can be used to control the flow of thermal currents, the weak vdW bonds conversely lead to characteristically low thermal boundary conductance (TBC) with 3D contacts, particularly with insulating substrates where the TBC is phonon dominated \cite{PopMRSBulletin2012}. Since the primary pathway for heat removal in 2D devices has been shown to be across the 2D/3D interface (depicted in Fig. \ref{fig:introschem}(a) and through the 3D substrate \cite{YasaeiAMI17}, the interfacial TBC can pose a bottleneck for heat removal \cite{Yasaei2DMats2017}. The relatively poor TBC of 2D/3D interfaces compared to covalently-bonded 3D/3D interfaces is a pressing issue for the development and implementation of nanoelectronic devices based on 2D materials. One path forward is to select the optimal substrate pairing for each 2D material that maximizes TBC, which requires a thorough understanding of the physical properties that impact heat transfer between the 2D material and substrate.


Concurrent to the development of 2D-material platforms, the application of machine learning (ML) in material science over the last two decades has led to new insights and advancements \cite{HautierCM2010, OlivaresEES2011, MuellerRCC2016, JainJMR2016, RamprasadNPJ2017, RyanJACS2018, GraserChemMat2018, SchmidtNPJ2019, SadatJAP2020}. More recently, ML has been used to investigate the thermal properties of materials \cite{CarretePRX2014, WanNL2019}, including the TBC between 3D/3D \cite{ZhanSciRep2017, WuNpj2019} and 2D/2D systems \cite{YangNanoscale2018}. In the former, the authors collected experimental data for the TBC of 3D/3D interfaces which they use to compare ML predictions to acoustic and diffuse mismatch models (AMM and DMM) and found that Gaussian process and support vector regression can outperform the AMM or DMM when predicting experimental data \cite{ZhanSciRep2017}. Such a comparison was facilitated by the abundance of experimental data on 3D/3D interfaces as compared to 2D/2D or 2D/3D interfaces. 
 
There have been several successful efforts to experimentally measure 2D/3D thermal boundary conductance (TBC) \cite{ZChenAPL09, YangJAP2014, YalonAMI2017, YasaeiAMI17, Yasaei2DMats2017, YasaeiAM2018} as well as to provide explanations of the underlying physical dynamics through theoretical modeling based on Green's functions \cite{PerssonJPCM10,PerssonJPCM11,OngPRB16,OngPRB17}, molecular dynamics (MD) \cite{OngJAP18,SuryavanshiJAP2019}, and the Boltzmann transport equation \cite{CorreaNT17, Foss2DMat2019,FossNanotechnol2021,FossPRM2020}. However, the vast majority of these studies focus on a few interface pairs involving graphene (Gr), hBN, and transition metal dichalcogenides (TMDs) on SiO$_2$, while measurements of interfaces involving other substrates (AlO$_x$, AlN, and diamond) \cite{Yasaei2DMats2017} were primarily done with Gr as the 2D layer. Although these studies represent some of the first investigations and theoretical frameworks of 2D/3D TBC, the literature lacks the breadth of 2D/3D TBC data (both experimental and theoretical) that is required to train most ML algorithms. To date, there are fewer than 50 experimentally measured or theoretically calculated values for different interface pairs \cite{YueaNanoRev15, Yasaei2DMats2017, Foss2DMat2019}, which is insufficient to train a predictive model. There is a need to expand the available data so that accurate TBC prediction and methodical materials selection can be accomplished. 

In this work, we use our first-principles-driven and experimentally validated phonon Boltzmann transport model to compute the thermal boundary conductance of 156 different 2D/3D interface pairs on a range of vdW spring coupling constants $K_a=[0.4,8]$ N.m$^{-1}$. The details of the 2D/3D vdW TBC model are given in Section S1 of the supporting information. Fig. \ref{fig:introschem} shows the general workflow from data generation, collection, to machine learning analysis. Our substrates include many widely used and technologically relevant substrates, such as SiO$_2$, high conductivity materials like GaN and diamond, alongside newly proposed ones of interest for 2D integration, such as HfO$_2$ and CaF$_2$ \cite{IllarionovNatCom2020}. This greatly expands upon available data in the literature and highlights several 2D and 3D materials as being candidates for ultra-high or -low TBC. We then use our model to calculate the TBC of each interface pair on a range of vdW coupling constants resulting in 5460 observations. This enables the training of ML algorithms to develop streamlined predictive models that are computationally cheap, numerically accurate to DFT+pBTE predictions, and demonstrate strong transferability. 

We demonstrate that when using an exhaustive list of material descriptors, excellent model fitness showing $R^2>0.9$ and root-mean-square error (RMSE) $\epsilon < 15$ M\tbcu~ can be achieved for numerous ML algorithms: decision-tree, neural network, and Gaussian process regression. We then use decision-tree regression to perform sensitivity analysis and select-down a list of material descriptors that are the most influential towards predicting the TBC. Sensitivity analysis determines that the vdW spring coupling constant, 2D thermal conductivity, ZA phonon bandwidth, ZA phonon resonance gap, and the frequency of the first van Hove singularity (for crystal substrates) or Boson peak (for amorphous substrates) are the most influential material descriptors. For ultra-high TBC, our calculations highlight BAs/PMMA (polymethyl methacrylate), hBN/CaF$_2$, and BAs/SiO$_2$ as having the highest TBC amongst interfaces calculated here with values of 178, 138, and 115 M\tbcu, respectively. Of these, only hBN/CaF$_2$ involves a crystalline substrate while the others are on amorphous substrates. Conversely, diamond, crystalline Al$_2$O$_3$, and 6H-SiC show TBC less than 10 M\tbcu~ for most interface pairs, owing to the mismatch between soft flexural 2D phonons and stiff crystalline substrates, with WSe$_2$/diamond being the lowest at 0.18 M\tbcu. Overall, we find that 2D/3D TBC can span 3 orders of magnitude.  

\begin{figure*}[t]
	\includegraphics[width=.9\textwidth]{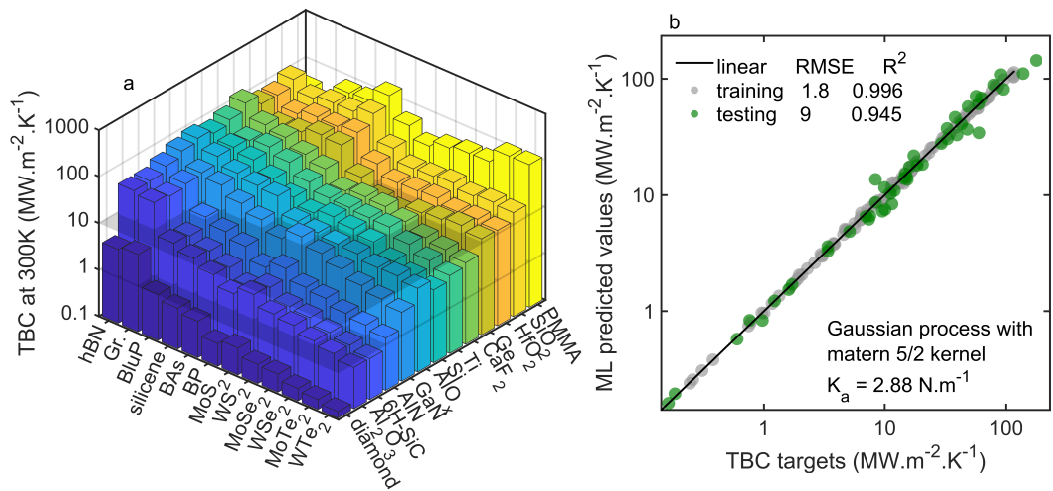}
	\caption{Room-temperature TBC (a) of each 2D/3D pair at a constant value of spring coupling constant (K$_a$).  The fitness of a Gaussian process regression model is assessed in (b) where a testing RMSE of $\le$10 M\tbcu~is achieved against DFT+pBTE simulations of TBC for 156 2D/3D interface pairs at a constant vdW spring coupling constant (K$_a$ =2.88 N.m$^{-1}$).}\label{fig:barplot}
\end{figure*}
The room-temperature TBC of each interface pair calculated with our model at a vdW coupling of $K_a=2.88$ N.m$^{-1}$ is shown in Fig. \ref{fig:barplot}(a). Graphene and hBN stand out as the best performing 2D materials for interface thermal transport with TBC $>$ 10 M\tbcu~ for most pairings except for diamond. However, graphene is a semi-metal and hBN is electrically insulating, and so neither can be used as a channel material in field-effect-transistor devices where semiconductors with an appropriate bandgap (0.4-2 eV) are required. Out of the semiconducting 2D materials studied here (BAs, BluP, BP, and TMDs), we find that BluP, BAs, and BP outperform the heavier TMDs. As for substrate (3D) materials, we find that PMMA, SiO$_2$, HfO$_2$, and Ge are the four best performing substrate materials for 2D/3D TBC. However, fcc calcium fluoride (CaF$_2$), which has been shown to form well-defined vdW interfaces with 2D materials \cite{KomaAS1990}, also shows very high TBC (138 M\tbcu) with hBN. 


We first trained machine learning models using the room-temperature TBC data of the 156 interface pairs at only a single value of vdW spring coupling constant. We found that Gaussian process and support vector regression could predict the TBC with good predictive accuracy RMSE $\le$ 11 M\tbcu~(shown in Fig. S1 in the Supporting Information). Details on how material descriptors were compiled and machine learning model parameters can be found in section S2 of the Supporting Information. The model fitness for Gaussian process regression with a matern-5/2 kernel function is shown in Fig. \ref{fig:barplot}(b) where excellent predictive accuracy is seen. Although sufficient predictive accuracy can be achieved on a small set of 156 datapoints, the resulting ML model with a single $K_a$ does not quantify the relationship between $K_a$ and TBC, which we previously found to be strongly related \cite{Foss2DMat2019}. Furthermore, $K_a$ can vary significantly due to unmitigated sample-to-sample variations in interface preparation \cite{ZChenAPL09, PopNL2017, YasaeiAMI17}. Processing steps, such as annealing, have been shown to impact the morphology of the substrate surface \cite{ParkJAP21}, which affects vdW spring coupling through average atomic separation between the 2D layer and the substrate surface. First-principles predictions of $K_a$ are hampered by the incommensurabilty of 2D layers and crystalline substrates or aperiodicity of amorphous substrates, which we elaborate further in Supplemental Information section S1.

To overcome these limitations, we compute the TBC for each unique interface over a range of vdW coupling constants spanning [0.4,8] N.m$^{-1}$ resulting in a total of 5460 observations. By using an adjustable $K_a$ one can span a range of potentially achievable values of TBC provided the required adhesion/coupling can be experimentally achieved. Further, training the ML models on a range of values of $K_a$ results in a more versatile model that allows one to subsequently predict a range of possible TBC values for any other arbitrary interface. The complete TBC dataset along with machine learning scripts that can be used to reproduce all ML results can be found at our lab website, given in the Data Availability. Using the full list of 16 material descriptors, we trained numerous regression models and found the best performing algorithms to be Gaussian process (GPR with a matern-5/2 kernel) and neural network (NN) regression. Both of these algorithms display excellent accuracy on the testing set with RMSE $<$ 5 M\tbcu~ and $R^2 >$ 0.99. However, such ML models are often considered complex \textit{black-box} models that can heavily distort and obscure input-output relations \cite{WanNL2019,WangCM2020}.

\begin{figure}[t]
	\centering
	\hspace*{-0.65cm}\includegraphics[width=1.2\columnwidth]{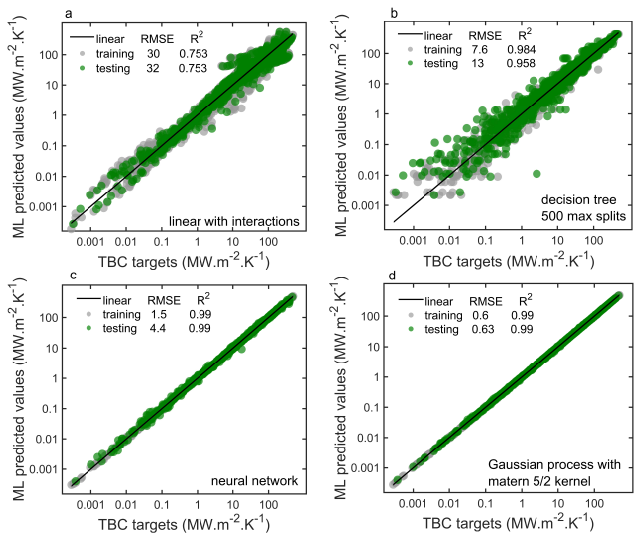}
	\caption{Machine learning model fitness for (a) linear (with interactions), (b) binary decision-tree, (c) neural network, and (d) Gaussian process regression algorithm is shown. The black solid line represents an exact linear fit to the target data, while the grey and green circles represent the training and testing/validation data, respectively.}\label{fig:fit1}
\end{figure}

Therefore, we also look at the performance of simpler ML models such as linear and decision tree regression which are known to provide better interpretability when compared to more sophisticated ML models (NN, GPR). The best performing of these simpler ML algorithms are linear with interactions (LRi) and binary decision-tree regression (DTR). We show the fitness results for training these four ML algorithms (a-LRi, b-DTR, c-NN, and d-GPR) in Fig. \ref{fig:fit1} where grey and green circles represent training and testing predictions, respectively. The RMSE and coefficient of determination $R^2$ remains under $\epsilon<15$ M\tbcu~and above 0.95, respectively, for all algorithms except for linear with interactions. The RMSE is dominated by interface pairings with larger TBC values (greater than 10 M\tbcu) across all models. This is exampled best by the LRi ML model in Fig. \ref{fig:fit1}(a) where larger deviations in predicted values greater than 10 M\tbcu~compared to other ML algorithms can be seen in the top-right corner of the fitness plot. 

	

Next, we perform a sensitivity analysis to determine which material descriptors are the most impactful to model training. We found that decision tree models preserved model fitness better than a linear with interactions model, particularly when removing descriptors. Therefore we chose to perform our full sensitivity analysis and feature selection using an ensemble of decision trees which are also well known for their high interpretability \cite{SagiInf2021} and use in feature-selection \cite{PolewkoPS2020}. It is not clear if performing the sensitivity analysis with a linear model would produce the same 5-descriptor set, however we suspect there to be overlapping terms. Our sensitivity analysis is performed by iteratively removing material descriptors and determining model fitness (RMSE and $R^2$) each time. We then find the descriptor that when removed improves model fitness, and remove it. We then repeat the process having removed that descriptor until the model does not improve upon removing any descriptors. Further details on sensitivity analysis can be found in section S8 of the Supporting Information. This results in a list of the material descriptors that are the most impactful on model accuracy -- these descriptors are the vdW coupling constant ($K_a$), 2D thermal conductivity ($\kappa_{2D}$), ZA phonon bandwidth ($BW_{ZA}$), ZA phonon resonance gap ($\omega_0$), and the location of first substrate vDOS peak ($\omega_{3D}$). 

\begin{figure*}[tbh!]
	\centering
	\includegraphics[width=0.9\textwidth]{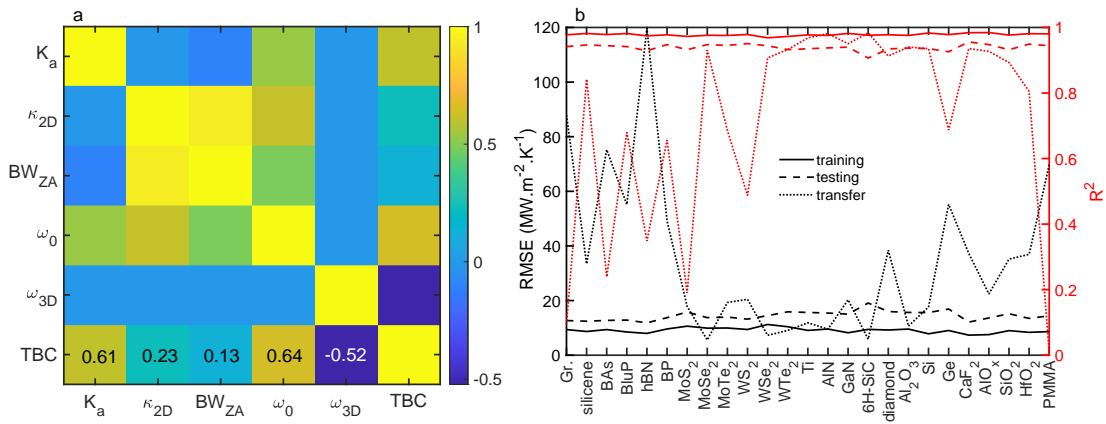}
	\caption{The Pearson correlation coefficients between the \textit{leading} 5 descriptors and the TBC is shown in (a). Using the reduced set of 5 descriptors to train a binary decision-tree algorithm we then analyze the transferability of the model to pairings not used in the training. Root-mean-square-error (RMSE) and the coefficient of determination ($R^2$) for a binary decision-tree algorithm tested on materials it has not seen listed along the x-axis is shown in (b).}\label{fig:SA}
\end{figure*}

The Pearson correlation coefficients of the most influential material descriptors identified by sensitivity analysis are shown in \ref{fig:SA}(a). Pearson correlation coefficients show the linearity between descriptors and the TBC where values close to $\pm$1 represent strong positive/negative correlation and values close to zero represent low correlation. It can be seen that the ZA resonant frequency gap $\omega_0$ is the most correlated descriptor to the TBC. This is due to two reasons: (1) $\omega_0$ depends on the vdW spring coupling constant $K_a$, which also shows high correlation, and the 2D unit cell mass as $\omega_0 = \sqrt{K_a/m_{2D}}$, and (2) $\omega_0$ impacts the shape of the 2D vDOS at low-energies \cite{PerssonEPL10,TalebJPCM16,CorreaNT17}. The latter affects the substrate scattering rate (seen in Section S1 of the Supporting Information) which has a $1/\omega^2$ dependence. On the other hand, $\omega_{3D}$ has a strong negative correlation and describes the frequency of the first peak of the susbtrate vDOS -- for crystalline materials this is the first van Hove singularity and for amorphous substrates this is the boson peak \cite{EtienneJNCS2002, ShintaniNMat08, YangPRB2021}. These descriptors highlight the dependence of the TBC on the shape of the vDOS on each side of the interface. 

A complete set of material descriptors used for training ML models is shown in Table 1 of the Supporting Information Section S2 and the corresponding Pearson correlation map for all descriptors can be found in Section S4. We note that the 5 most-correlated descriptors may not be the 5-most impactful descriptors as determined by sensitivity analysis. The reason is because the Pearson correlation coefficients only highlight linear relationships \cite{WilhelmBook2008} through a normalized covariance. The sensitivity analysis we performed is aimed at minimizing the model RMSE while removing descriptors. It therefore captures non-linear relationships between descriptors and the TBC. Moreover, despite the substrate Debye temperature ($\theta_{3D}$) showing high correlation to the TBC in the Pearson correlation map (seen in Supporting Information Section S4), it is not selected by sensitivity analysis. This is likely due to non-linear dependencies as well as the high intercorrelation between $\theta_{3D}$ and other descriptors such as $\omega_{3D}$ which is selected by sensitivity analysis.

We next determine the \textit{transferability} of decision tree algorithms. Transferability is the ability of the ML model to transfer its displayed predictive accuracy to data it has not been trained on (i.e., data the model has not seen). To do this, we iteratively remove 2D and 3D materials from the training and validation cycles. We then perform subsequent predictions for those interfaces involving the excluded material. The decision tree model is retrained each time with the missing material information, this ensures that each subsequent \textit{transferred} prediction is a true \textit{blind} prediction. The model fitness parameters for a decision tree algorithm resulting from the removal of each material studied here is shown in Fig. \ref{fig:SA}(b). We see accurate predictions (RMSE less than 20-30 M\tbcu~ and an $R^2>0.75$ in most cases) on the TBC of removed materials such as silicene, TMDs, nitrides, oxides, and fcc-structures (e.g., Si and CaF$_2$) except for Ge. 
	
Some outliers are graphene (Gr), boron arsenide (BAs), boron nitride (hBN), germanium (Ge), and PMMA which represent the lightest 2D (having low $m_{2D}$), heavy crystalline substrates, and lightest amorphous material in our dataset. This merely illustrates the expected limitation that the ML model cannot predict the TBC for interface pairs that lie outside the extrema of material descriptor values. To explain why there is a larger $R^2$ for hBN as compared to graphene or PMMA, yet even larger RMSE for hBN as compared to the same, we point the reader to the transferability analysis fitness results in Supporting Information section S9. For hBN, the transferred predictions (orange circles) occupy the top-right corner of the fitness curve, where the TBC is the largest. Therefore, the RMSE for deviations in that region are very large leading to a large RMSE for hBN despite it having a better transferred $R^2$ than graphene or PMMA. Other examples of the model fitness of transferred predictions of removed materials are shown in Section S9 of the Supporting Information.

Lastly, we look at the model fitness of each four ML models (used in Fig. \ref{fig:fit1}) when using just the five leading material descriptors. We see a clear degredation in the predictive accuracy of LRi. However, the remaining ML models retain high predictive accuracy, showing only a modest increase in the RMSE and $R^2$ values in the NN and GPR cases. This shows that even complex machine learning algorithms, such as NN and GPR, can still retain reasonable predictive accuracy when using a reduced set of only five material descriptors. The reduced descriptor set can then be useful for initial searches or pre-screens for potential candidate pairings with desirable properties, to be followed-up by with more accurate methods, such as the full descriptor set, first-principles modeling, or experimental measurements.
		
	\begin{figure}
		\centering
		\hspace*{-0.65cm}\includegraphics[width=1.2\columnwidth]{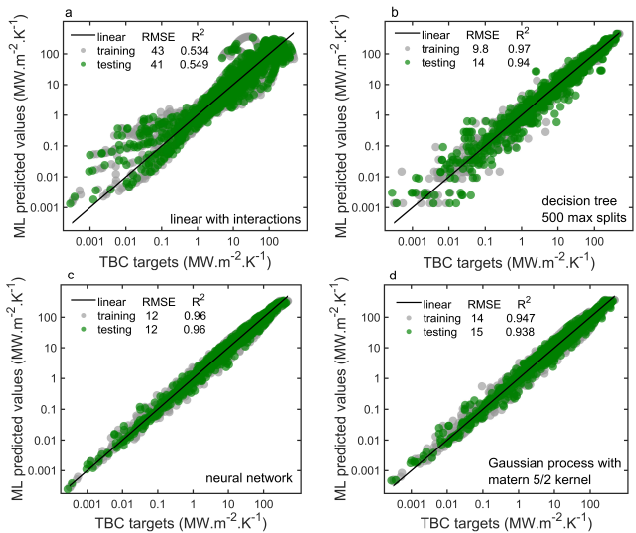}
		\caption{The fitness of (a) linear, (b) binary decision-tree, (c) neural network, and (d) Gaussian process regression algorithms trained using just the 5 \textit{leading} descriptors are shown.}\label{fig:fit2}
	\end{figure} 
	
	
The number of observations needed to train a machine learning model and obtain good predictive accuracy is an important metric when data is limited. Therefore, it is advantageous to estimate how much data (and thereby effort) is required before machine learning algorithms can be expected to return accurate predictions when trained on experimental data. We find that the number of observations needed to train fine/dense decision tree algorithms is roughly 1500 (shown in Fig. S3 in the Supporting Information). 
	
In conclusion, we have computed the room-temperature thermal boundary conductance (TBC) of 156 2D/3D pairings over a range of van der Waals coupling constants $K_a=[0.4,8]~N.m^{-1}$ to produce a data set of 5460 TBC values. The generation of this data enabled the application of machine learning algorithms to study the thermal interface problem of 2D/3D TBC. Our data is based on rigorous first-principles-driven phonon Boltzmann transport calculations and includes a wide range of material and interface properties as material descriptors in the training of machine learning algorithms. We use sensitivity analysis coupled with the high interpretability of decision tree regression to identify a list of 5 material descriptors (from a total list of 16) that most heavily impact ML model fitness. Those leading material descriptors are the vdW spring coupling constant $K_a$, 2D thermal conductivity $\kappa_{2D}$, ZA phonon bandwidth $BW_{ZA}$, the ZA phonon resonance gap $\omega_0$, and the frequency of the first van Hove singularity or Boson peak $\omega_{3D}$. For 2D materials beyond graphene and hBN, we find that BluP, BAs, and BP outperform the heavier TMDs. As for substrate (3D) materials, we identify PMMA, along with SiO$_2$, HfO$_2$, and Ge as the best-performing substrate materials for 2D/3D TBC with a TBC of 177 M\tbcu~in BAs/PMMA. Further, we discover that CaF$_2$ is a crystalline substrate that exhibits very high TBC (138 M\tbcu) with hBN at room-temperature while WSe$_2$/diamond is the lowest with only 0.18 M\tbcu. Our work will enable the selection of optimal substrate for any 2D material and pave the way toward 2D nano- and opto-electronic devices with improved thermal management. 
	
\section {Author Contributions}
C.F. and Z.A. conceived the work together. Z.A. implemented the TBC code while C.F. implemented the ML training and performed first principles and ML calculations. C.F. created all the figures. Both authors contributed to writing the manuscript. 

\section{Supplementary Material}
A document containing Supplemental Material is provided with this article. It contains additional figures in support of the manuscript as well as further details of the thermal boundary conductance model and the sensitivity analysis algorithm we employed to select the 5 descriptors. 

\section{Acknowledgments}
This work was supported by the National Science Foundation (NSF) program Computational and Data-Enabled Science and Engineering (CDS\&E) grant 1902352 to Z.A.

\section{Data Availability Statement}
All data resulting from this work and used in this paper is available on reasonable request from the authors. Dataset used to train the ML model will also be hosted publicly on the corresponding author's website https://nanoenergy.mse.utah.edu for download. 

\section{Conflict of Interest}
The Authors declare no Conflicting Financial or Non-Financial Interests.


%

\end{document}